\documentclass[twocolumn,showpacs,amssymb]{revtex4}

\usepackage{graphicx}
\usepackage{dcolumn}
\usepackage{bm}   

\begin{document}

\title{Large Fluctuations, Classical Activation, Quantum Tunneling, and Phase Transitions}

\author{D.L. Stein}
\email{dls@physics.arizona.edu}

\affiliation{Departments of Physics and Mathematics, University of Arizona,
Tucson, AZ 85721 USA}

\date{\today}

\begin{abstract}

We study two broad classes of physically dissimilar problems, each
corresponding to stochastically driven escape from a potential well.  The
first class, often used to model noise-induced order parameter reversal,
comprises Ginzburg-Landau-type field theories defined on finite intervals,
perturbed by thermal or other classical spatiotemporal noise.  The second
class comprises systems in which a single degree of freedom is perturbed by
both thermal and quantum noise.  Each class possesses a transition in its
escape behavior, at a critical value of interval length and temperature,
respectively.  It is shown that there exists a mapping from one class of
problems to the other, and that their respective transitions can be
understood within a unified theoretical context.  We consider two
applications within the first class: thermally induced breakup of
monovalent metallic nanowires, and stochastic reversal of magnetization in
thin ferromagnetic annuli.  Finally, we explore the depth of the analogy
between the two classes of problems, and discuss to what extent each case
exhibits the characteristic signs of critical behavior at a sharp
second-order phase transition.

\end{abstract}


\maketitle


\section{Introduction}
\label{sec:intro}

\vskip -.075in

Noise-induced escape from a locally stable state governs a wide range of
physical phenomena~\cite{GarciaOjalvo99}.  In spatially extended classical
systems, where the noise is typically, but not necessarily, of thermal
origin, these phenomena include homogeneous nucleation of one phase inside
another~\cite{Langer69}, micromagnetic domain
reversal~\cite{Neel49,Brown63}, pattern formation in non-equilibrium
systems~\cite{CH93}, and many others.  The concepts and techniques employed
are formally identical to those used in tunneling problems in quantum field
theories, including among others the `decay of the false
vacuum'~\cite{CC79}, anomalous particle production~\cite{tHooft76}, and
macroscopic quantum tunneling in a dc SQUID~\cite{MSIB94}.

Recently a new kind of `phase transition', in the classical activation
behavior of spatially extended systems perturbed by weak spatiotemporal
noise, has been uncovered and analyzed~\cite{MS01,MS03,Stein04}.  The
existence of a similar crossover, from thermal activation to quantum
tunneling in systems as simple as a single degree of freedom, has long been
known~\cite{Goldanskii59,Affleck81,Wolynes81,CL81,LO83,GW84,RHF85,Chudnovsky92,KT97,GB97,FY99}.
Here we will review the stochastic escape behaviors of both types of
systems, explore the depth of the analogy between them, discuss to what
extent the crossover can be thought of as a type of second-order phase
transition, and show how the purely classical transition, occurring as
system length is varied, can shed light on the crossover from classical
activation to quantum tunneling as temperature is lowered.  But let's begin
with a purely classical problem.

\vskip -.05in

\section{Model}
\label{sec:model}

\vskip -.05in

Consider an extended system described by a classical field $\phi(z,t)$
defined on the spatial interval $[-L/2,L/2]$, subject to the potential
$V(\phi)$ and perturbed by spatiotemporal white noise.  Its time evolution
is governed by the stochastic Ginzburg-Landau equation
\begin{equation}
\label{eq:GL}
\partial_t\phi=\partial_{zz}\phi - \partial_\phi V(\phi) +
\sqrt{2T}\xi(z,t)\, ,
\end{equation}
where all dimensional quantities have been scaled out.  The first term on
the RHS arises from a field `stiffness', i.e., an energy penalty for
spatial variation of the field.  The system is stochastically perturbed by
additive white noise $\xi(z,t)$, satisfying
$\langle\xi(z_1,t_1)\xi(z_2,t_2)\rangle=\delta(z_1-z_2)\delta(t_1-t_2)$.
We consider only weak noise, i.e., the noise magnitude $T$ is small
compared to all other energy scales in the problem (formally, our analysis
will be asymptotically valid in the $T\to 0$ limit).  The frictional
coefficient has been set to one, so when $T$ is temperature
Eq.~(\ref{eq:GL}) obeys the fluctuation-dissipation relation.

The deterministic, or zero-noise, dynamics can be written as the variation
of an action ${\cal H}$ with the field $\phi$:
\begin{equation}
\label{eq:variation}
\dot\phi=-\delta{\cal H}/\delta\phi\, ,
\end{equation}
and because the dynamics is gradient, the action ${\cal H}$ is simply an
energy functional 
\begin{equation}
\label{eq:energy}
{\cal H}[\phi]\equiv \int_{-L/2}^{L/2}dz\ \left[{1\over
2}(\partial_z\phi)^2 +V(\phi)\right] \, .
\end{equation}

\begin{figure}
\begin{center}
\begin{tabular}{c}
\includegraphics[height=5.0cm]{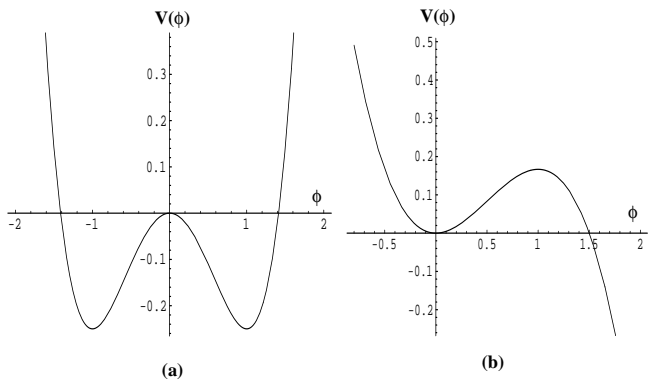}
\end{tabular}
\end{center}
\caption[The activation energy.]{\label{fig:potentials} Potentials used in
discussion: (a) bistable quartic (Eq.~(\ref{eq:quartic})) and (b)
monostable cubic (Eq.~(\ref{eq:cubic})).}
\end{figure}

We will mostly consider relatively simple potentials, in particular the 
symmetric bistable quartic potential
\begin{equation}
\label{eq:quartic}
V_s(\phi)=-{1\over 2}\phi^2+{1\over 4}\phi^4\, ,
\end{equation}
and the asymmetric monostable cubic potential
\begin{equation}
\label{eq:cubic}
V_a(\phi)={1\over 2}\phi^2-{1\over 3}\phi^3\, .
\end{equation}
Both potentials are shown in Fig.~\ref{fig:potentials}.

In the weak-noise ($T\to 0$) limit, the classical activation rate for a
transition out of a stable well is
\begin{equation}
\label{eq:Kramers}
\Gamma\sim \Gamma_0\exp(-\Delta E/T)\, ,
\end{equation}
where $\Delta E$ is the activation barrier (to be defined below) and
$\Gamma_0$ the rate prefactor, determined by fluctuations about the most
probable escape path.  When the top of the barrier is locally quadratic,
the prefactor $\Gamma_0$ is {\it independent\/} of temperature.  In such
circumstances the escape rate~(\ref{eq:Kramers}) is said to be of the
Arrhenius-van't Hoff (or often simply Arrhenius) form.

\section{The Infinite-Domain Case}
\label{sec:infinite}

Let's start by considering the $L=\infty$ case, which was fully worked out
for the classical nucleation problem by Langer~\cite{Langer69} and for the
quantum tunneling problem by Callan and Coleman~\cite{CC79}.  (See also
Schulman~\cite{Schulman81} for a good pedagogic treatment.)  For
illustrative purposes, we will use the bistable quartic
potential~(\ref{eq:quartic}).

The states of interest --- i.e., the stable and saddle field configurations
--- are time-independent solutions of the zero-noise Ginzburg-Landau
equation $\dot\phi=-\delta{\cal H}/\delta\phi$.  They are therefore extrema
of the action so in the current case satisfy the nonlinear differential
equation
\begin{equation}
\label{eq:ELquartic}
\phi''=-\phi+\phi^3\, .
\end{equation}
Its spatially uniform solutions are just the stable states $\phi(z)=\pm 1$
and the local maximum $\phi=0$.  But our main interest is in the
soliton-like pair (or `bounce'~\cite{CC79})
\begin{equation}
\label{eq:tanh}
\phi(z)=\pm\tanh\Big[(z-z_0)/\sqrt{2}\Big]
\end{equation}
where $z_0$ is a constant whose presence denotes the fact that the `domain
wall', i.e.~the spatially varying piece of~(\ref{eq:tanh}) that separates
the two stable states, can nucleate anywhere on the line.  
\begin{figure}
\begin{center}
\begin{tabular}{c}
\includegraphics[height=6.25cm]{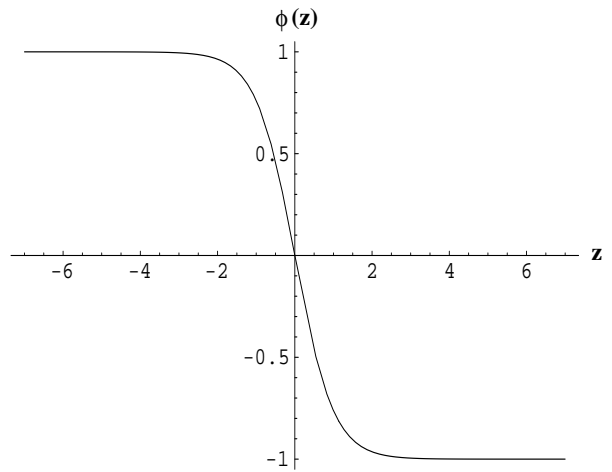}
\end{tabular}
\end{center}
\caption[The activation energy.]{\label{fig:tanh} The `bounce' described by
Eq.~(\ref{eq:tanh}) with $z_0=0$.  For clarity, only one of the symmetric
pair is shown.}
\end{figure}

Assuming for the moment that~(\ref{eq:tanh}) is the saddle configuration
(this will be justified below), the activation energy $\Delta E=E[\phi_{\rm
saddle}]-E[\phi_{\rm stable}]$ in the Kramers rate
formula~(\ref{eq:Kramers}) can be computed, giving
\begin{eqnarray}
\label{eq:infdeltaE}
\Delta E=\int_{-\infty}^\infty dz\ \Big[{1\over 2}(\partial_z\phi_{\rm
saddle})^2+V(\phi_{\rm
saddle})\Big]\nonumber\\-\int_{-\infty}^\infty dz\ \Big[{1\over
2}(\partial_z\phi_{\rm stable})^2+V(\phi_{\rm stable})\Big]={2\sqrt{2}\over
3}\, ,
\end{eqnarray}
which is essentially the energy of a single domain wall.

In order to compute the prefactor $\Gamma_0$ of the Kramers
rate~(\ref{eq:Kramers}), we need to examine fluctuations about the optimal
escape path, in particular in the vicinity of the stable and saddle field
configurations.  The procedure for doing this is described in detail
elsewhere~(see, for example,~\cite{Schulman81,HTB90,Stein04}), and will be
simply summarized here.

Let ${\bf\varphi}_s$ denote the stable state (in this case, the uniform
$\pm 1$ state), and let ${\bf\varphi}_t$ denote the transition (saddle)
state (here given by Eq.~(\ref{eq:tanh})).  Consider a small perturbation
${\bf\eta}$ about the stable state, i.e.,
${\bf\varphi}={\bf\varphi}_s+{\bf\eta}$.  Then to leading order
$\dot{\bf\eta}=-{\bf\Lambda}_s{\bf\eta}$, where ${\bf\Lambda}_s$ is the
linearized zero-noise dynamical operator governing the time evolution of
fluctuations about ${\bf\varphi}_s$.  Similarly ${\bf\Lambda}_t$ is the
linearized zero-noise dynamical operator for ${\bf\varphi}_t$.

We next diagonalize the linear time-evolution operators, by decomposing
fluctuations about the stable and transition states into normal modes,
which are eigenfunctions $\eta_i$ of the cooresponding operators:
\begin{equation}
\label{eq:ev}
{\bf\Lambda}_b\eta_i=\lambda_i\eta_i\, ,
\end{equation}
where $b=s,t$.  An eigenfunction with positive eigenvalue $\lambda_i>0$ is
a stable mode; one with $\lambda_i<0$ is unstable.  A stable (or
metastable) state therefore has all $\lambda_i>0$; a saddle state has a
single $\lambda_i<0$.  Its corresponding eigenfunction denotes the unstable
direction (in function space) in the vicinity of the saddle, leading either
back to the initial stable configuration or out of the well entirely.  With
the potential~(\ref{eq:quartic}), the eigenvalue equation becomes
\begin{equation}
\label{eq:phib}
\dot\eta = -{\bf\Lambda}_b\eta \equiv -\left[-{d^2}/{dz^2} + (-1 +
  3\phi_b^2)\right]\eta\, .
\end{equation}

In most cases the barrier is locally quadratic: all eigenvalues are nonzero
and we're left with an infinite set of decoupled quadratic fluctuations
about the stable and saddle states.  Then~\cite{HTB90,MM89}
\begin{equation}
\label{eq:gamma0}
\Gamma_0 = 
\frac1{2\pi}
\sqrt{\left|\frac{\det{\bf\Lambda}_s}{\det{\bf\Lambda}_t}\right|}
\,\,\left|\lambda_{t,0}\right|,
\end{equation}
where $\lambda_{t,0}$ is the only negative eigenvalue of~${\bf\Lambda}_t$.
In general, the determinants in~(\ref{eq:gamma0}) separately diverge: they
are products of an infinite number of eigenvalues with magnitude greater
than one.  However, their {\it ratio\/}, which can be interpreted as the
limit of a product of individual eigenvalue quotients, is finite.

There is a technical difficulty that has not yet been addressed, namely the
existence of a {\it soft collective mode\/} corresponding to the
arbitrariness of $z_0$ in~(\ref{eq:tanh}): the domain wall (or `instanton')
can nucleate anywhere.  The resulting translational symmetry implies a
zero-eigenvalue mode.  Its removal can be achieved with the McKane-Tarlie
regularization procedure~\cite{MT95} (see also~\cite{KC98}) for functional
determinants.  This will not be discussed further here, except to note that
its presence results, among other things, in a non-Arrhenius
(noise-dependent) prefactor that scales with the length.  The overall
result for the prefactor {\it per unit length\/} is then
\begin{equation}
\label{eq:prefinf}
\Gamma_0/L=\Bigl({4\sqrt{6}\over\pi}\Bigr)\Bigl({2\over\sqrt{\pi T}}\Bigr)\, ,
\end{equation}
where the first term on the RHS follows from the computation of
Eq.~(\ref{eq:gamma0}) with the zero eigenvalue removed via the
McKane-Tarlie procedure, and the second term (divided through by length
$L$) gives the contribution of the zero eigenvalue, i.e., the effect of the
translational symmetry of the bounce.

\section{The Finite-Domain Case}
\label{sec:finite}

The differences between the infinite and finite-line cases are not only
quantitative; there are important qualitative differences that are also
relevant, in an entirely different context, to the
classical~$\leftrightarrow$~quantum crossover~(Sec.~\ref{sec:crossover}).
The most striking of these differences is a sharp change in activation
behavior as interval length is varied; moreover, this change exhibits
characteristics of a second-order phase
transition~\cite{MS01,MS03,Stein04}, but only in a strictly asymptotic
sense.

Because we are now working on an interval of finite length, we need to
specify boundary conditions.  With the exception of the zero-energy mode
that accompanies only translation-invariant (such as periodic) boundary
conditions, different choices of boundary conditions lead only to minor
quantitative differences.  To avoid the (minor) complication of the zero
mode altogether, we will employ Neumann boundary conditions:
$\partial\phi/\partial L|_{-L/2}=\partial\phi/\partial L|_{L/2}=0$.  With
this choice $\phi_{\rm stable}=\pm 1$, as before.  We will continue to use
the notation of Sec.~\ref{sec:infinite}, where $\phi_s$ ($\phi_t$) refers
to the stable (transition) state.

For the symmetric $\phi^4$ potential with Neumann boundary conditions, the
change in activation behavior arises from a {\it bifurcation\/} of the
transition state, from a uniform configuration below a critical length
$L_c$ to a pair of degenerate, spatially varying `bounce' configurations
above $L_c$.  More precisely, the saddle states are
\begin{equation}
\label{eq:uniform}
\phi_t=0 
\end{equation}
when $L< L_c$ and
\begin{equation}
\label{eq:bounce}
\phi_t=\pm\sqrt{2m\over 1+m}{\rm sn}({x\over\sqrt{m+1}}\mid m)\, ,
\end{equation}
when $L\ge L_c$, where ${\rm sn}(\cdot\mid m)$ is the Jacobi elliptic $\rm
sn$ function with parameter~$0\le m\le 1$.  Its quarter-period is given
by~${\bf K}(m)$, the complete elliptic integral of the first
kind~\cite{Abramowitz65}, which is a monotonically increasing function
of~$m$.  As~$m\to 0^+$, ${\bf K}(m)$ decreases to~$\pi/2$, and ${\rm
sn}(\cdot\mid m)\to\sin(\cdot)$.  In this limit the saddle state smoothly
degenerates to the $\phi=0$ configuration.  As $m\to 1^-$, the
quarter-period increases to infinity (with a logarithmic divergence), and
${\rm sn}(\cdot\mid m)\to\tanh(\cdot)$, the (nonperiodic) single-kink
sigmoidal function.  The Langer/Callan--Coleman bounce
solution~(\ref{eq:tanh}) is thereby recovered as $L\to\infty$.

\begin{figure}
\begin{center}
\begin{tabular}{c}
\includegraphics[height=6.25cm]{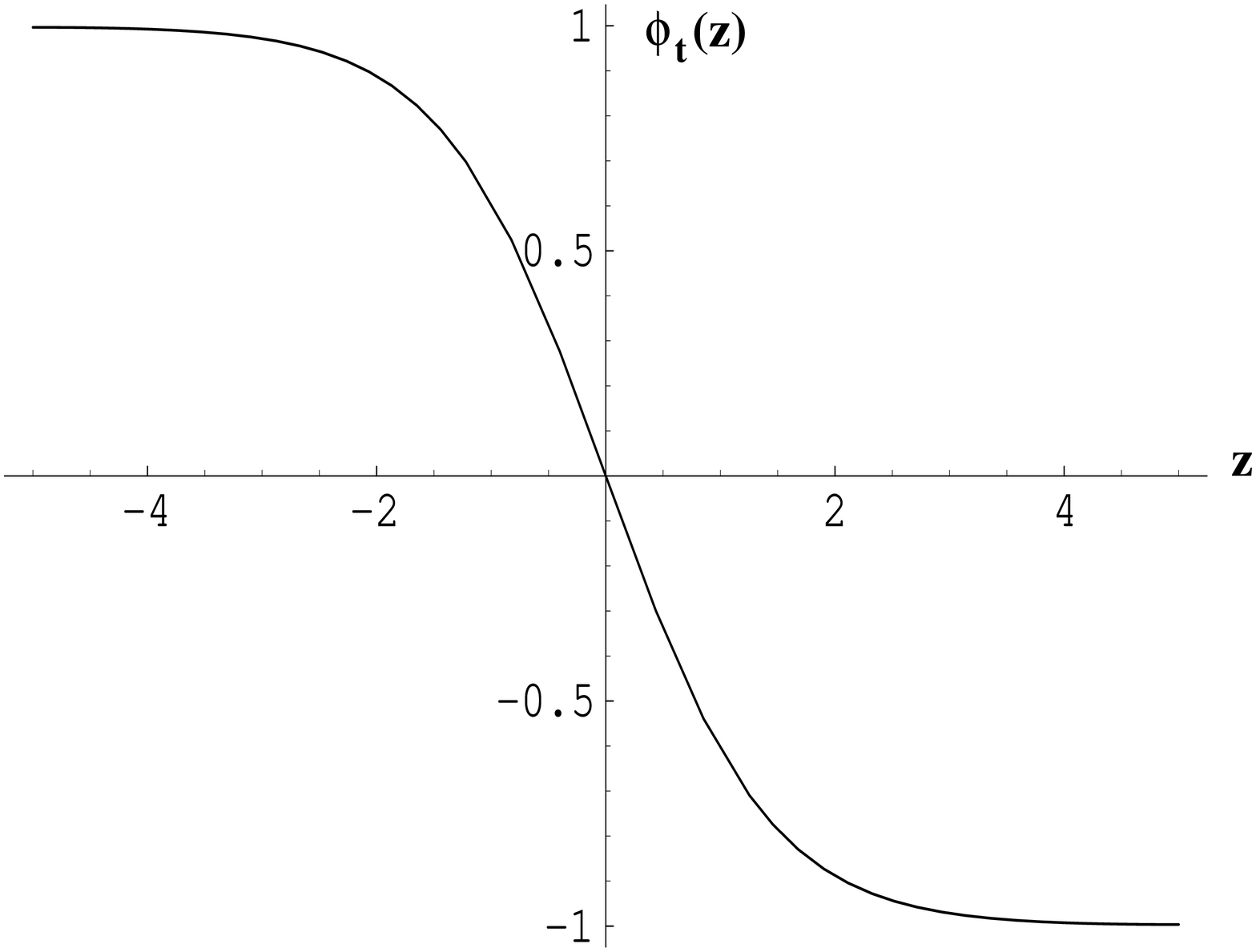}
\end{tabular}
\end{center}
\caption[The activation energy.]{\label{fig:neumannsol} The transition
 state $\phi_t(z)$ for $L=10$ (corresponding to $m=0.986$) described by
 Eq.~(\ref{eq:bounce}). As in Fig.~\ref{fig:tanh}, only one of the
 symmetric pair is shown.}
\end{figure}

The value of~$m$ in~(\ref{eq:bounce}) is determined by the interval length
$L$ and the Neumann boundary conditions, which require that
\begin{equation}
\label{eq:bc}
L/\sqrt{m+1}=2{\bf K}(m)\, .
\end{equation}
The critical length is determined by~(\ref{eq:bc}) when $m=0$; that is,
$L_c=\pi$.  As previously noted, $m\to 1$ corresponds to $L\to\infty$, and
the activation energy smoothly approaches the asymptotic value of
$2\sqrt2/3$.  The transition state for an intermediate value of $m$,
corresponding to $L=10$, is shown in Fig.~\ref{fig:neumannsol}.

\begin{figure}
\begin{center}
\begin{tabular}{c}
\includegraphics[height=6.25cm]{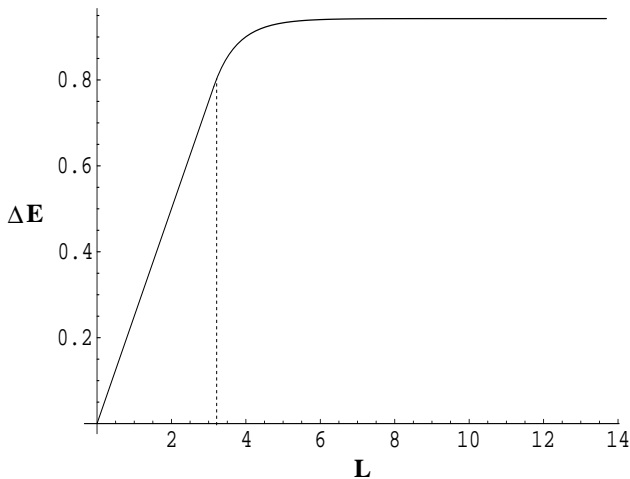}
\end{tabular}
\end{center}
\caption[The activation energy.]{\label{fig:acten} The activation energy
$\Delta E$ as a function of the interval length~$L$, for the potential
given by Eq.~(\ref{eq:quartic}) (with all coefficients set equal to one)
and Neumann boundary conditions.  The dashed line indicates the critical
interval length $L_c=\pi$ at which the saddle state bifurcation takes
place.}
\end{figure}

The activation energy $\Delta E$ can be solved in closed form for all $L>L_c$ (below
$L_c$, it is simply $L/4$):
\begin{equation}
\label{eq:acten}
\Delta E={1\over3(1+m)^{3/2}}\Bigl[4(1+m){\bf E}(m)-{1\over
2}(1-m)(3m+5){\bf K}(m)\Bigr]\, ,
\end{equation}
with ${\bf E}(m)$ the complete elliptic integral of the second
kind~\cite{Abramowitz65}.  The activation energy as a function of $L$ is
shown in Fig.~\ref{fig:acten}.  Note that the curve of $\Delta E$ vs.~$L$
and its first derivative are both continuous at $L_c$; the second
derivative, however, is discontinuous, as might be expected of a
second-order-like phase transition.

A more profound manifestation of critical behavior at $L_c$ is exhibited by
the rate prefactor $\Gamma_0$. When $L<L_c=\pi$, both the stable and saddle
states are spatially uniform: $\phi_{\rm stable}=\pm1$ and $\phi_{\rm
saddle}=0$.  This greatly simplifies the computation of the associated
eigenvalues.  

It is immaterial which of the two (degenerate) stable states is used; the
eigenvalue spectrum is the same at both because of the symmetry under
$\phi\mapsto-\phi$.  Linearizing around either stable state yields the
operator
\begin{equation}
\label{eq:phis}
{\bf\Lambda}_s=-d^2/dz^2+2\, ,
\end{equation}
and similarly
\begin{equation}
\label{eq:phiu}
{\bf\Lambda}_t=-d^2/dz^2-1\, .
\end{equation}
The eigenvalue spectrum of ${\bf\Lambda}_s$ with Neumann boundary
conditions is
\begin{equation}
\label{eq:stablespectrum}
\lambda_{s,n}=2+{\pi^2n^2\over L^2}\qquad\qquad\qquad n=0,1,2\ldots\,
\end{equation}
The eigenvalue spectrum of ${\bf\Lambda}_t$ is similarly
\begin{equation}
\label{eq:unstablespectrum}
\lambda_{t,n}=-1+{\pi^2n^2\over L^2}\qquad\qquad\qquad n=0,1,2\ldots\,.
\end{equation}
As required, all eigenvalues of ${\bf\Lambda}_s$ are positive, while
${\bf\Lambda}_t$ has a single negative eigenvalue $\lambda_{t,0}=-1$.  Its
eigenfunction, which is spatially uniform, is the direction in
configuration space along which the optimal escape trajectory
approaches~$\phi_t$.

Putting everything together, we find the Neumann-case rate prefactor when
$L<L_c$ to be
\begin{eqnarray}
\label{eq:g0-}
\Gamma_0&=&{1\over2\pi}\,\sqrt{{\prod_{n=0}^\infty\bigl(2+{\pi^2n^2\over
L^2}\bigr)\over
\left|\prod_{n=0}^\infty\big(-1+{\pi^2n^2\over L^2}\big)\right|}}\nonumber \\
&=&
\frac{1}{2^{3/4}\pi}\sqrt{\frac{\sinh(\sqrt2L)}{\sin L}}\, .
\end{eqnarray}
As $L\to L_c^-(=\pi^-)$, $\Gamma_0\sim (L_c-L)^{-1/2}$.  This divergence
has a simple physical interpretation: the optimal escape trajectory becomes
transversally unstable, in the direction defined by the eigenmode $\eta_1$,
as the critical length is approached.  Mathematically the divergence is
caused by $\lambda_{t,1}\to 0^+$ as $L\to L_c^-$.

When $L>L_c$, there are two transition states, namely the nonuniform bounce
configurations $\pm\phi_t$ given by~(\ref{eq:bounce}).  The associated
linearized evolution operator, computed from~(\ref{eq:phib}), is
\begin{equation}
\label{eq:phiuop}
{\bf\Lambda}_t= -{d^2\over dz^2}-1+{6m\over 1+m}\ {\rm
sn}^2\left({x\over\sqrt{m+1}}\Big|\ m\right)\, .
\end{equation}

\begin{figure}
\begin{center}
\begin{tabular}{c}
\includegraphics[height=6.25cm]{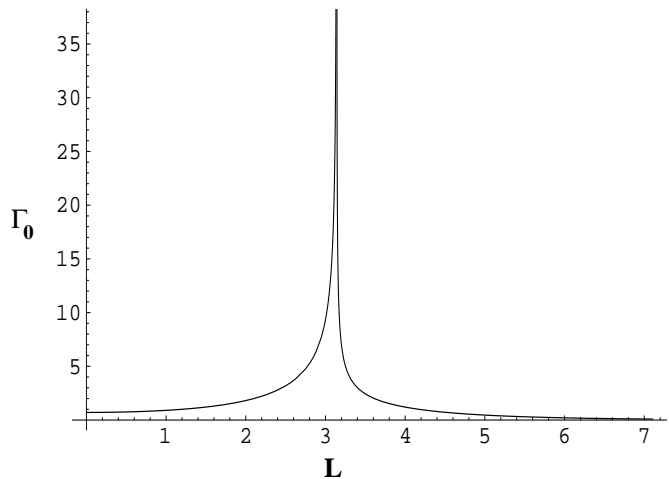}
\end{tabular}
\end{center}
\caption[The prefactor.]{\label{fig:prefactor} Rate prefactor $\Gamma_0$ for
the quartic potential with Neumann boundary conditions, showing the
power-law divergence of the prefactor as $L\to L_c^\pm$.}
\end{figure}

Calculation of the associated determinant quotient, and the single unstable
eigenvalue $\lambda_{t,0}$, is described in detail in~\cite{MS03}, to which
the interested reader is referred.  The Neumann-case rate prefactor when
$L>L_c$ is
\begin{eqnarray}
\label{eq:prefabove}
\Gamma_0= {1\over \pi} \left|1-{2\over 1+m}\sqrt{m^2-m+1}\right|\nonumber\\
\times\sqrt{{{\sinh(\sqrt{2}L)}\over{\sqrt{2}\left|(1-m){\bf K}(m)-(1+m){\bf E}(m)\right|}}}
\, .
\end{eqnarray}
As~$m\to 0^+$ ($L\to L_c^+$), $\Gamma_0\sim(L-L_c)^{-1/2}$.  The prefactor
over the entire range of $L$ is shown in Fig.~\ref{fig:prefactor}.

The divergence at $L_c$ is striking, but requires interpretation.  We defer
further discussion to Sec.~\ref{sec:discussion}, and turn now to physical
applications of the methods and results presented in this section.

\section{Two Applications}
\label{sec:applications}

We have established a transition in activation behavior as system length is
varied; but is it, and more generally the theory presented above, relevant
to actual systems?  In this section we briefly discuss two applications.

\subsection{Lifetimes of monovalent metallic nanowires}
\label{subsec:nanowires}

Metallic nanowires are cylindrically-shaped incompressible electron fluids
with diameters of order tens of atoms and with lengths hundreds to
thousands of atoms.  They are stabilized by quantum shell
effects~\cite{KSGG01,ZKS03,UG03} but at nonzero temperatures are only
metastable, with breakup probably due to thermal
fluctuations~\cite{YYR99,YYR00,YRY01}.  We have proposed~\cite{BSS04} a
self-consistent continuum approach to studying the lifetimes of monovalent
metallic nanowires, with a large-deviation-induced `collapse' modelled
through a stochastic Ginzburg-Landau field theory, of the kind discussed
above.  Our theory provides good quantitative agreement with available data
on nanowire lifetimes, and accounts for the observed difference in
stability between alkali and noble metal nanowires.

\begin{figure}
\begin{center}
\begin{tabular}{c}
\includegraphics[height=6.25cm]{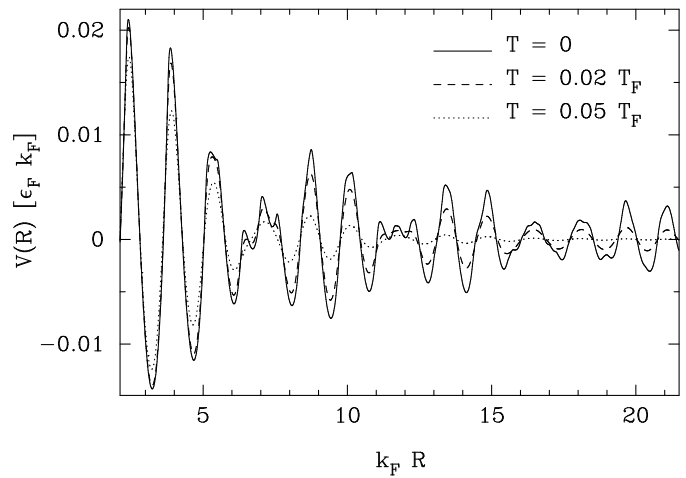}
\end{tabular}
\end{center}
\caption{\label{fig:nanowire} Electron-shell potential $V_{\rm shell}(R)$
at three temperatures, computed from the free-electron model of
Ref.~\cite{SBB97}.}
\end{figure}

We treat a nanowire as a cylinder of length~$L$ and
radius~$R(z)=R_0+\phi(z)$, with $z\in[-L/2,L/2]$.  Radius fluctuations are
governed by the stochastic Ginzburg-Landau equation~(\ref{eq:GL}),
where~$V(\phi)$ arises from the zero-temperature electron-shell potential
of~Fig.~\ref{fig:nanowire}.

Details of the calculations appear in~\cite{BSS04}; here we simply
summarize the results, which appear in Table~\ref{tab:lifetime}.

\begin{table*}
\caption{The lifetime $\tau$ (in seconds) for various cylindrical sodium
and gold nanowires at temperatures from 75K to 200K.  Here $G$ is the
electrical conductance of the wire in units of $G_0=2e^2/h$, $L_c$ is
the critical length, and $\Delta E_\infty$ is the activation energy for an
infinitely long wire.  From~Ref.~\cite{BSS04}.  }
\label{tab:lifetime}
\vskip .1in
\begin{tabular}{|c||c|c|c|c|c||c|c|c|c|c|}
\hline
\rule[-1.5ex]{0pt}{4.5ex} & \multicolumn{5}{c||}{Na} & \multicolumn{5}{c|}{Au} \\
\cline{2-11}
\rule[-1.5ex]{0pt}{4.5ex} $G$ & $L_c $ & $\Delta E_\infty $ &  \multicolumn{3}{c||}{$\tau$ [s]}  & $L_c $ & $\Delta E_\infty $ &  \multicolumn{3}{c|}{$\tau$ [s]} \\
\cline{4-6}\cline{9-11}
\rule[-1.5ex]{0pt}{4.5ex} [$G_0$] & [nm] & [meV] & $75\;$K & $100\;$K & $150\;$K & [nm] & [meV] & $100\;$K & $150\;$K & $200\;$K \\
\hline\hline
\rule[0ex]{0pt}{3.ex}%
3 & 1.4 &  210 & 30 &  $9\times10^{-3}$ &  $3\times10^{-6}$ & 1.5 &  470 &  $10^{11}$ &  $2\times10^{3}$ & 0.2 \\
6 & 2.6 &  170 & 0.06 &  $9\times10^{-5}$ &  $10^{-7}$ & 3.0 &  310 &  $10^{3}$ &  $7\times10^{-3}$ &  $2\times10^{-5}$ \\
17 & 3.1 &  230 &   500 & 0.08 &  $10^{-5}$ & 3.4 &  470 &  $9\times10^{10}$ &  $10^{3}$ & 0.2 \\
23 & 3.7 &  190 & 4 &  $2\times10^{-3}$ &  $10^{-6}$ & 4.1 &  390 &  $10^{7}$ & 3 &  $2\times10^{-3}$ \\
42 & 4.3 &  210 & 50 & 0.01 &  $4\times10^{-6}$ & 4.8 &  440 &  $3\times10^{9}$ &   100 & 0.03 \\
51 & 4.5 &  150 &  $7\times10^{-3}$ &  $2\times10^{-5}$ &  $5\times10^{-8}$ & 4.9 &  320 &  $2\times10^{3}$ & 0.01 &  $3\times10^{-5}$ \\
\rule[-1.5ex]{0pt}{3.ex}%
96 & 5.8 &  200 & 5 &  $2\times10^{-3}$ &  $10^{-6}$ & 6.3 &  440 &  $8\times10^{9}$ &   300 & 0.05 \\
\hline
\end{tabular}
\end{table*}

The lifetimes tabulated for sodium nanowires in Table \ref{tab:lifetime}
exhibit a rapid decrease in the temperature interval between 75K and 100K.
These lifetimes correlate well with the observed temperature dependence of
conductance histograms for sodium nanowires~\cite{YYR99,YYR00,YRY01}.  A
comparison of the lifetimes of sodium and gold nanowires listed in
Table~\ref{tab:lifetime} indicates that gold nanowires are much more
stable.  In our model this arises from the difference in surface tension:
$\sigma_{\rm Au}= 5.9\, \sigma_{\rm Na}$, and is consistent with the
observation that noble metal nanowires are much more stable than alkali
metal nanowires.

There is also an important prediction contained in
Table~\ref{tab:lifetime}, namely that nanowire lifetimes, which exhibit
significant variations from one conductance plateau to another, do not vary
systematically as a function of radius.  It can be seen from
Table~\ref{tab:lifetime} that the activation barriers vary by only about
30\% from one plateau to another, and that a wire with a conductance of $96
G_0$ has essentially the same lifetime as that with a conductance of $3
G_0$.  In this sense, the activation barrier variation exhibits {\em
universal mesoscopic fluctuations}: in any conductance interval, there are
very short-lived wires (not shown in Table \ref{tab:lifetime}) with very
small activation barriers, while the longest-lived wires have activation
barriers of a universal size:
\begin{equation}
\label{eq:universal}
0\,< \,\Delta E_\infty\, \lesssim \, 0.7 
\left(\frac{\hbar^2 \sigma}{m_e}\right)^{1/2}\!\!\!\!\!,
\label{eq:Delta_E}
\end{equation}
The derivation of~(\ref{eq:universal}) will be presented
elsewhere~\cite{BSSinprep}.

At present, lifetimes of wires shorter than $L_c$ have not been
systematically studied.  We expect that over the next several years the
technology will improve to the point where this will become possible, and
our predictions for a transition in both activation barriers (from barriers
essentially independent of length to those varying linearly with length, as
in Fig.~\ref{fig:acten}) and prefactors (Fig.~\ref{fig:prefactor}) can be
tested.

\subsection{Magnetic reversal in nanomagnets}
\label{subsec:nanomagnets}

The dynamics of magnetization reversal in submicron-sized, single-domain
particles and thin films is important for information storage and other
magnetoelectronic applications.  This problem can be treated with the
methods used throughout this paper, but with a more complicated equation of
motion than~(\ref{eq:GL}).  The magnetization dynamics is governed by the
Landau-Lifschitz-Gilbert equation~\cite{LDE80} perturbed by weak
spatiotemporal noise:
\begin{equation}
\label{eq:LLG}
\partial_t{\bf M}=-\gamma[{\bf M}\times{\bf H}_{\rm
eff}]+(\alpha/{M_0})[{\bf M}\times\partial_t{\bf M}]\, ,
\end{equation}
where $M_0$ is the magnitude of the magnetization ${\bf M}$, $\alpha$ the
damping constant, and $\gamma>0$ the gyromagnetic ratio.  The effective
field ${\bf H}_{\rm eff}=-\delta E/\delta{\bf M}$ is the variational
derivative of the total energy $E$, which is given by (with free space
permeability $\mu_0=1$):
\begin{eqnarray}
\label{eq:energy0}
E[{\bf M}({\bf x})]&=&\lambda^2\int_\Omega d^3x |\nabla{\bf
M}|^2+\int_{{\bf R}^3}d^3x|\nabla U|^2\nonumber\\ &-&\int_\Omega d^3x {\bf
H}_{\rm e}\cdot {\bf M}\, ,
\end{eqnarray}
where $\Omega$ is the region occupied by the ferromagnet, $\lambda$ is the
exchange length, and $U$ (defined over all space) satisfies
$\nabla\cdot(\nabla U+{\bf M})=0$.  The first term on the RHS
of~(\ref{eq:energy0}) is the bending energy, the second the magnetostatic
energy, and the last the Zeeman energy.  We take ${\bf H}_{\rm
e}\parallel\hat\theta$, using cylindrical coordinates $(\hat
r,\hat\theta,\hat z)$.  The magnetostatic energy is nonlocal and gives rise
to shape anisotropies (for `soft' magnetic materials, like fcc Co or
permalloy, crystalline anisotropies are negligible).  The out-of-plane
anisotropy energy is especially strong, and forces ${\bf M}$ to lie in the
plane.

The presence of the nonlocal magnetostatic term complicates analysis.
However, the quasi-$2D$ nature of the problem allows a significant
simplification, as shown by Kohn and Slastikov~\cite{KS04}. Their
asymptotic scaling analysis applies to the present problem when both the
aspect ratio $t/R$ (ring thickness divided by ring mean radius), and
$\lambda/R$, are sufficiently small.  Because of the high energy cost of
variations in $M_0$, and because the geometry under consideration admits
nonsingular solutions for the vector field ${\bf M}$, $M_0$ can be taken to
be fixed.  The result for the total magnetic energy is then~\cite{MSK04}:
\begin{equation}
\label{eq:magenergy}
{\cal E}=\int_0^{\ell/2}ds \Big[({\partial\phi\over\partial
s})^2+\sin^2\phi-2h\cos\phi\Big]\, ,
\end{equation}
where $\ell$ is related to ring circumference, $h$ is the external magnetic
field magnitude, and $\phi$ is the local angle between the in-plane
magnetization vector and the local field direction.
In~(\ref{eq:magenergy}) energy, length, and field are all dimensionless,
normalized by a corresponding characteristic quantity determined by various
ring parameters.

Eq.~(\ref{eq:LLG}) and the variational equation ${\bf H}_{\rm eff}=-\delta
E/\delta{\bf M}$ yield a nonlinear differential equation that must be
satisfied by any time-independent solution:
\begin{equation}
\label{eq:EL}
d^2\phi/ds^2= \sin\phi\cos\phi + h\sin\phi\, .
\end{equation}

There are three `constant' ($\phi$ independent of $\theta$) but nonuniform
(${\bf m}={\bf M}/M_0$ varies with position) solutions for $0\le h<1$: the
stable state $\phi=0$ (${\bf m}=\hat\theta$); the metastable state
$\phi=\pi$ (${\bf m}=-\hat\theta$), and a pair of degenerate unstable
states $\phi=\cos^{-1}(-h)$, which constitute the saddle for a range of
$(\ell,h)$. The $\phi=0,\pi$ solutions are degenerate when $h=0$, and the
$\phi=\pi$ solution becomes unstable at $h=1$.

We have also found a nonconstant `bounce' solution of~(\ref{eq:EL}), which
is the saddle for the remaining range of $(\ell,h)$.  It is
\begin{equation}
\label{eq:instanton}
\phi(s,m)=2\cot^{-1}\Bigl[{\rm dn}\Big({s-s_0\over\delta}\Big|m\Big){{\rm
sn}({\cal R}|m)\over{\rm cn}({\cal R}|m)}\Bigr]\, ,
\end{equation}
where dn$(\cdot|m)$, sn$(\cdot|m)$, and cn$(\cdot|m)$ are the Jacobi
elliptic functions with parameter $m$~\cite{Abramowitz65}, $s_0$ is a
constant, and ${\cal R}$ and $\delta$ are given by
\begin{eqnarray}
\label{eq:aux}
{\rm sn}^2({\cal R}|m)&=&1/m-h/2\nonumber\\
&-&(1/2m)\sqrt{m^2h^2+4(1-m)}\\
\delta^2&=&{m^2\over 2-(m+\sqrt{m^2h^2+4(1-m)})}\, .
\end{eqnarray}
Imposition of the periodic boundary condition yields a relation between
$\ell$ and $m$:
\begin{equation}
\label{eq:pbc}
\ell=2{\bf K}(m)\delta\, .
\end{equation}
As $m\to 0$, dn$(x|0)\to1$, and the bounce solution reduces to the constant
state $\phi=\cos^{-1}(-h)$.  The critical length and field where this
occurs are related by
\begin{equation}
\label{eq:crit}
\ell_c=\pi\delta_c=2\pi/\sqrt{1-h_c^2}\, .
\end{equation}

\begin{figure}
\begin{center}
\begin{tabular}{c}
\includegraphics[width=0.48\textwidth]{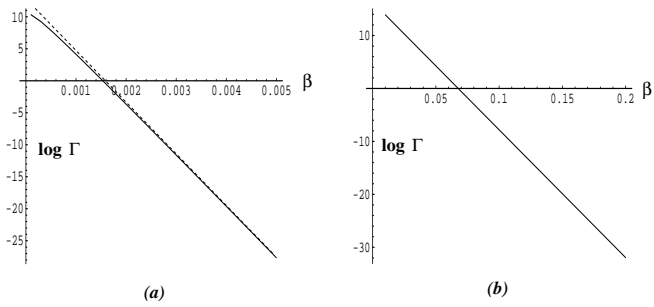}
\end{tabular}
\end{center}
\caption{Total switching rate (in units of $s^{-1}$) vs.~inverse
temperature~$\beta$ (in units of $^\circ K^{-1}$), at fields of (a) 52.5 mT
(nonconstant saddle) and (b) 72.5 mT (constant saddle). Parameters used are
$k=.01$, $l=.05$, $R=200$ nm, $R_1=180$ nm, $R_2=220$ nm, $M_0=8\times
10^5$ A/m (permalloy), $\alpha=.01$, and $\gamma=1.7\times
10^{11}T^{-1}s^{-1}$. Deviation of low-field switching rate in (a) from
dashed line signals non-Arrhenius behavior. (From Ref.~\cite{MSK04}.)}
\label{fig:switching}
\end{figure}

What is interesting here is that the transition is governed by {\it two\/}
parameters: not only the length, but also the magnitude of the externally
applied magnetic field.  While the former cannot be varied continuously,
the latter can, allowing for the first time a detailed experimental probe
of the transition.  Fig.~\ref{fig:switching} shows theoretical predictions
of the magnetization switching rate at two different field strengths for a
ring of fixed circumference.

\section{Crossover from Thermal Activation to Quantum Tunneling}
\label{sec:crossover}

Only purely classical activation processes have been considered so far, but
the transition in activation behavior as interval length is increased has
an interesting parallel with the crossover from classical activation to
quantum tunneling as temperature is lowered.  The correspondence can be
made explicit by mapping system length $L$ in the former case to
temperature $T$ in the latter.

The partition function $Z$ at inverse temperature $\beta$ can be written as
a Euclidean path integral over trajectories $q(\tau)$, where $\tau=it$ is
imaginary time, with each path weighted by its Euclidean action
$S_E$~\cite{Feynman72}:
\begin{equation}
\label{eq:path}
Z=\int\ {\cal D}[q(\tau)]\exp\{-S_E[q(\tau)]/\hbar\}\, .
\end{equation}
The integral in~(\ref{eq:path}) runs over all paths periodic in imaginary
time, with period~$\hbar\beta$.

Eq.~(\ref{eq:path}) is derived from the usual definition of the partition
function $Z={\rm Tr}\ \{\exp[-\beta{\cal H}]\}$, where ${\cal H}$ includes
both the system {\it and\/} its environment.  Consequently, even if one is
dealing with the tunneling of only a single degree of freedom $q(\tau)$,
the effects of friction due to its coupling with the environment must be
included.  The proper treatment of the effects of damping on quantum
mechanical tunneling have been considered by a number of authors; see, for
example,~\cite{CL81,LO83,GW84,RHF85}.  Although friction strongly affects
the tunneling rate quantitatively, it will not be included here; this is
because our only aim is to present the connection between the transition in
thermal activation of (infinite-dimensional) classical fields as length is
varied, and the crossover from classical activation to quantum tunneling in
(one-dimensional) systems as temperature is varied.

An early treatment (that ignored dissipation) was given by
Goldanskii~\cite{Goldanskii59}. Setting the classical,
temperature-dependent Arrhenius factor $\Delta E/k_BT$ equal to the
zero-temperature quantum tunneling rate through a parabolic barrier, he
noted that the characteristic temperature $T_0$ for the quantum tunneling
$\leftrightarrow$ classical activation crossover was
\begin{equation}
\label{eq:Goldanskii}
T_0=\hbar\omega_c/(2\pi k_B)\, ,
\end{equation}
where $\omega_c$ is the characteristic frequency of the locally quadratic
barrier.  Although Goldanskii's approach led only to an estimate, his
formula~(\ref{eq:Goldanskii}) was quite accurate (in the absence of
dissipation), as the following more detailed
analysis~\cite{LO83,GW84,RHF85} will demonstrate (see
also~\cite{Affleck81,Wolynes81,Chudnovsky92,KT97,GB97,FY99}).

In what follows, we will use the asymmetric potential given
by~(\ref{eq:cubic}).  This allows us to consider only incoherent tunneling
processes: once the particle tunnels through the barrier, it escapes for
good.  A double-well potential such as~(\ref{eq:quartic}) requires
consideration of dissipative quantum coherence effects, for which a
real-time functional integral approach is better suited~(see, for
example,~\cite{BM82,CL84}). The Euclidean action for a particle of mass $M$
is therefore
\begin{eqnarray}
\label{eq:action}
S_E[q(\tau)]&=&\int_{-\beta\hbar/2}^{\beta\hbar/2}d\tau\ \Bigl\{{1\over 2}M{\dot
  q}^2(\tau)\\
&+&M\Big[{\omega_0^2\over 2}q(\tau)^2-{\lambda\over
  3}q(\tau)^3\Big]\Bigr\}+\hbox{(frictional terms)}\nonumber
\end{eqnarray}
where $\dot q$ denotes a derivative with respect to imaginary time $\tau$.
In the low-friction limit, extremal paths satisfy
\begin{equation}
\label{eq:qext}
-{\ddot q}(\tau) +\omega_0^2q(\tau)-\lambda q(\tau)^2=0\, ,
\end{equation}
with periodic boundary conditions $q(-\beta\hbar/2)=q(\beta\hbar/2)$.

Three extremal solutions are physically relevant.  The first is the uniform
$q(\tau)=0$ solution, which corresponds simply to the stable state at all
temperatures.  Of the other two, one is uniform:
\begin{equation}
\label{eq:qc+}
q_{c,+}(\tau)=\omega_0^2/\lambda
\end{equation}
and the other is a nonuniform bounce:
\begin{eqnarray}
\label{eq:qc-}
q_{c,-}(\tau,m)&=&{3\omega_0^2\over2\lambda}\nu(m)^2\ {\rm
  dn}^2\Bigl({\omega_0\nu(m)\over 2}(\tau-\tau_0)\mid m\Bigr)\nonumber\\
  &+&{\omega_0^2\over2\lambda}\Bigl[1-(2-m)\nu(m)^2\Bigr]\, ,
\end{eqnarray}
where as before $0\le m\le 1$ and $\nu(m)=(1-m+m^2)^{-1/4}$.  Imposition of
the periodic boundary condition relates the parameter $m$ to the inverse
temperature:
\begin{equation}
\label{eq:betabc}
\beta=4{\bf K}(m)/\hbar\omega_0\nu(m)\, .
\end{equation}
The constant $\tau_0$ in~(\ref{eq:qc-}) indicates the translational
symmetry in imaginary time arising from the periodic boundary conditions,
and corresponds to a zero mode as described in Sec.~\ref{sec:infinite}.

We can now make explicit the mapping to the thermal activation of classical
fields.  Here the inverse temperature $\beta$ plays the same role as system
length $L$ in Sec.~\ref{sec:finite}.  In particular, high temperature
corresponds to the small length regime.  For $\beta$ smaller than some
$\beta_c$, we would therefore expect the constant solution $q_{c,+}$ to be
the saddle.  And indeed it is.  The corresponding action is
\begin{equation}
\label{eq:highT}
S_E[q_{c,+}(\tau)]=M\omega_0^6\beta\hbar/(6\lambda^2)=\beta\hbar\Delta V\,
,
\end{equation}
where $\Delta V=M\omega_0^6/(6\lambda^2)$ is the potential energy
difference between the potential barrier top and bottom.  Consequently,
$\exp\{-S_E[q_{c,+}(\tau)]/\hbar\}=\exp[-\beta\Delta V]$, and the
classical Arrhenius factor is recovered.

The bounce solution $S_E[q_{c,-}(\tau)]$ is the saddle at large $\beta$.
As in Sec.~\ref{sec:finite}, the transition temperature $T_c$ is given
by~(\ref{eq:betabc}) when $m\to 0^+$.  We find that
$T_c=\hbar\omega_0/(2\pi k_B)$, exactly that found by Goldanskii
(Eq.~(\ref{eq:Goldanskii})), using much simpler arguments.  (For the
potential used in~(\ref{eq:action}), the curvatures at the top and bottom
of the well are equal.)

\begin{figure}
\begin{center}
\begin{tabular}{c}
\includegraphics[width=0.48\textwidth]{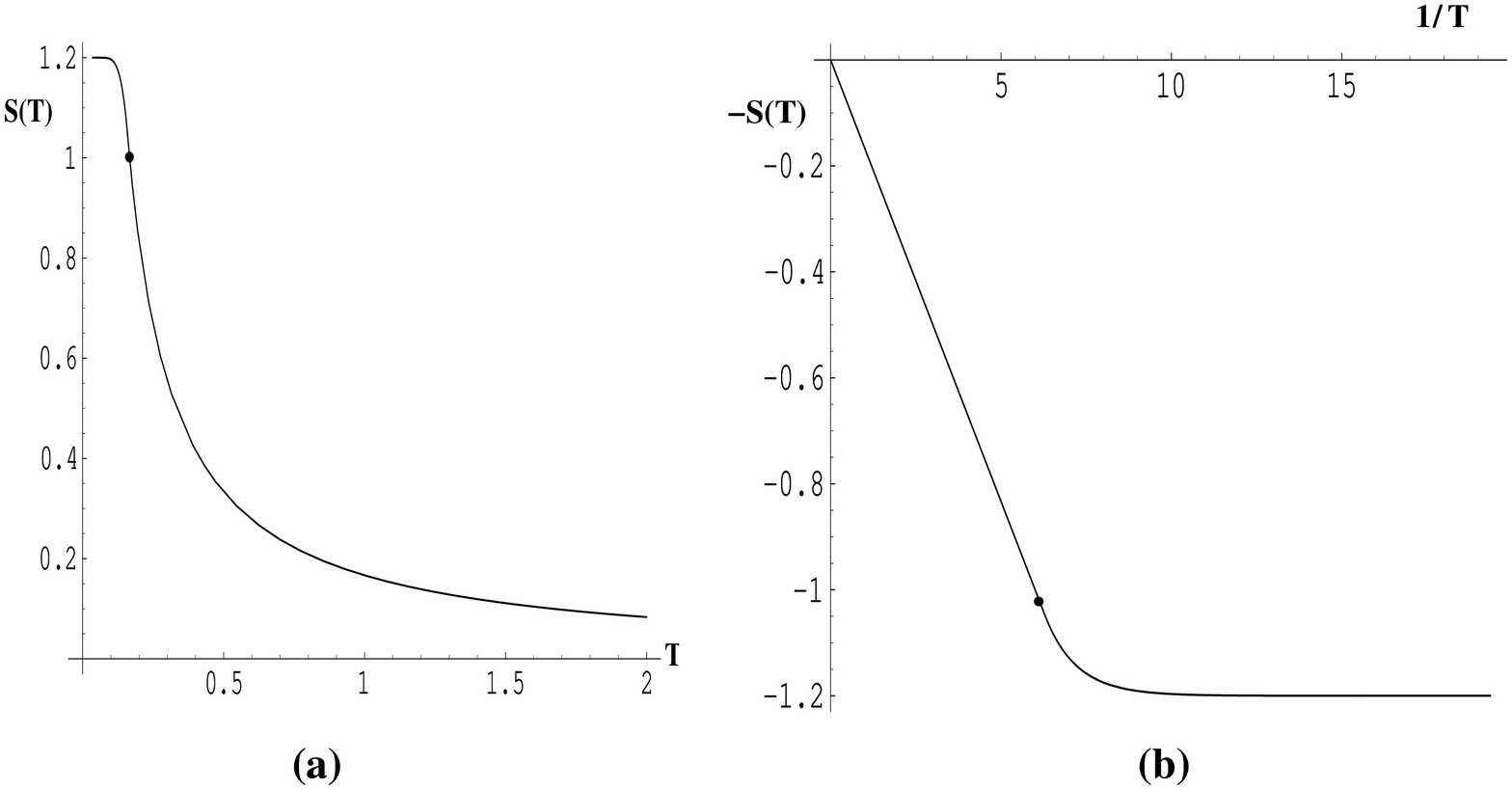}
\end{tabular}
\end{center}
\caption{(a) The Euclidean action $S_E[q_c(\tau)]$ at all temperatures.
(b) The logarithm of the leading-order escape rate, shown in an
Arrhenius-style plot.  In both graphs, $M=\omega_0=\lambda=\hbar=k_B=1$,
and the dot indicates the crossover.}
\label{fig:action}
\end{figure}

The action in the low-temperature region 
\begin{eqnarray}
\label{eq:lowT}
S_E[q_{c,-}]=3M\omega_0^5/(5\lambda^2(1-m+m^2)^{1/4})\nonumber\\
\times\Bigl[2{\bf E}(m) -[(2-m)(1-m)/(1-m+m^2)]{\bf K}(m)\Bigr]\nonumber\\
+(M\beta\hbar\omega_0^6/
12\lambda^2)\Bigl[1-3(2-m)/2\sqrt{1-m+m^2}\nonumber\\
+(2-m)^3/2(1-m+m^2)^{3/2}\Bigr]
\end{eqnarray}
is, as in the classical field case (cf.~Fig.~\ref{fig:acten}), continuous
and differentiable at all temperatures; but also as before, its second
derivative is discontinuous.  The action, along with the leading-order
escape rate, is shown in Fig.~\ref{fig:action}.

The well-known zero-temperature tunneling rate is recovered
from~(\ref{eq:lowT}) in the limit $m\to 1$.  Summarizing, we find that the
classical Arrhenius formula is recovered in the high-temperature limit and
the quantum tunneling formula is recovered in the zero-temperature limit:
\begin{eqnarray}
\label{eq:highlow}
\exp\Bigl[-S_E(q_{c,-})/\hbar\Bigr]=\left\{\begin{array}{r@{\quad\quad}l}
\exp[-\Delta V/k_BT]&T\to\infty\\
&\\
\exp[-36\Delta V/5\hbar\omega_0]&T\to 0\end{array}\right.
\end{eqnarray}
while in a narrow region about $T_0$ both contribute.

\section{Discussion}
\label{sec:discussion}

We have demonstrated that in problems involving noise-induced escape of a
classical field over a barrier, a type of second-order phase transition,
with what appears to be attendant critical phenomena (such as power-law
divergence of the rate prefactor) occurs as one or more external parameters
(length of the interval on which the field is defined, external magnetic
field if relevant, and so on) is varied.  We have also shown that there is
a mathematical mapping of this transition to the classical
activation~$\leftrightarrow$~quantum tunneling transition for a particle
escaping a simple potential well.  The mapping here involves identifying
interval length (and/or magnetic field, if appropriate) in the classical
field case to temperature in the classical~$\leftrightarrow$~quantum case.
To avoid confusion, it should be remembered that the asymptotically small
parameter is noise strength (typically, but not necessarily, temperature)
in the former problem, and Planck's constant --- {\it not\/} temperature
--- in the latter.

In this section we will address two questions that immediately come to
mind: To what extent can the transitions discussed be considered `real'
second-order phase transitions exhibiting critical phenomena; e.g., in the
sense that the fluctuations driving the transition occur on all scales?
Secondly, how deep is the correspondence between the classical field
transition and the quantum~$\leftrightarrow$~classical
crossover~\cite{Affleck81,Wolynes81,LO83,GW84,RHF85,Chudnovsky92,KT97,GB97,FY99}?

\subsection{Is the phase transition `real'?}
\label{subsec:real}

The short answer is: in a mathematically asymptotic sense yes, but from a
strictly (and more physically relevant) critical-phenomena-oriented
viewpoint, no.  Moreover, the observation of rate prefactor divergence
depends crucially on the order in which relevant parameters (temperature,
system length, and so on) are varied.  There are clearly different saddle
configurations, and therefore {\it qualitatively\/} different activation
behaviors, on either side of the transition. But the question we are
focusing on here is: what is happening very close to the critical
lengthscale?  This was extensively discussed in~\cite{Stein04}, and that
discussion will be expanded here.

Naively, the transition appears second-order in several respects: the
saddle solution is continuous (and even bifurcates in symmetric models such
as~(\ref{eq:quartic})), and the action is continuous and differentiable at
the transition point but has a discontinuous second derivative there.  But
perhaps most compelling, from the point of view of critical phenomena, is
the apparent power-law `divergence' of the rate prefactor shown in
Fig.~\ref{fig:prefactor}.  We therefore examine this in more detail, first
asking what it even means for the prefactor to `diverge'.

Of course, at no lengthscale is the actual prefactor infinite.  Consider
the analysis of the noisy symmetric Ginzburg-Landau model given in
Sec.~\ref{sec:finite}.  It is important to recall that the analysis of the
escape rate is, strictly speaking, valid only in an asymptotic sense as
$T\to 0$: our results apply {\it only\/} to temperatures sufficiently low
so that the escape rate is small.

What the formal divergence of the prefactor {\it does\/} mean is that the
escape behavior becomes increasingly anomalous as $L_c$ is approached, and
that it is {\it non-Arrhenius\/} exactly at $L_c$, where for all $T\to 0$
the prefactor is temperature-dependent, scaling as a negative power of $T$.
In the region close to $L_c$, the rate prefactor $\Gamma_0$ is anomalously
large, but still finite.  The formula~(\ref{eq:gamma0}) --- from which the
prefactor shown in Fig.~\ref{fig:prefactor} was computed --- is valid {\it
only\/} for $T$ sufficiently small so that the contributions from the
quadratic fluctuations about the relevant extremal state of ${\cal
H}[\phi]$ dominate the action.  So as long as all eigenvalues of
${\bf\Lambda}_t$ are nonzero (excluding, if translational symmetry is
present, the zero mode which may be extracted), Eq.~(\ref{eq:gamma0})
applies, but only within a temperature region driven to zero as $L\to L_c$.
The diminishing size of this region as $L\to L_c$ is controlled by the rate
of vanishing of the eigenvalue(s) of smallest magnitude.

The implication is that the prefactor behavior depends on whether $T\to 0$
at fixed $L$ near $L_c$, or whether $L$ increases (say) through $L_c$ at
fixed low $T$.  In the former case, one will recover
Fig.~\ref{fig:prefactor}.  If instead one fixes {\it temperature\/} at some
small but nonzero value, one should observe first a rising prefactor as $L$
approaches $L_c$, but at some $L$ (depending on $T$ through a type of
`Ginzburg criterion'; cf.~Fig.~9 in~\cite{Stein04}), the prefactor crosses
over to a non-Arrhenius (temperature-dependent) form.  As $L$ continues to
increase on the other side of $L_c$, the sequence of events is reversed.

The procedure of fixing $T$ and varying $L$ is in many instances the more
physical one.  In this case a correct analysis needs to include
higher-order (than quadratic) fluctuations about the transition state, as
was done, e.g., in~\cite{GW84}.  The next higher-order terms will (unless
prevented by symmetry) be nonzero.  The behavior remains anomalous,
however.  Suppose one starts to vary $L$ at fixed $T$, but as soon as
non-Arrhenius behavior is encountered, one fixes $L$ and then starts to
lower $T$. A subsequent transition from non-Arrhenius to Arrhenius behavior
will again be encountered: as $T$ is lowered, the prefactor~$\Gamma_0$
rises until it reaches the value shown in Fig.~\ref{fig:prefactor}; it
remains constant thereafter.

Lying behind this description is the relative magnitudes of the thermal
energy and that due to quadratic fluctuations in the direction of the
eigenmode $\eta_1$ with vanishing eigenvalue $\lambda_1$.  Non-Arrhenius
behavior should be seen at `intermediate' temperatures, where thermal
energy is large compared to that due to quadratic fluctuations along the
$\eta_1$ direction, but small compared to that arising from higher-order
fluctuations.  `Intermediate' here depends on $L$, whose closeness to $L_c$
determines the magnitude of $\lambda_1$.  Arrhenius behavior reappears in
the `low' temperature region where the thermal energy is small compared
even to that due to the quadratic fluctuations along $\eta_1$.  And because
$\lambda_1\to 0$ as $L\to L_c$, the crossover from non-Arrhenius to
Arrhenius behavior occurs at an increasingly lower temperature.  Exactly at
$L_c$, where $\lambda_1$ is exactly zero, the escape behavior is
non-Arrhenius for all $T\to 0$.

There's an alternative way of describing the situation: when viewed on
`normal' fluctuation lengthscales of $O(T^{1/2})$, field fluctuations along
the eigenmode direction $\eta_1$ appear to be diverging as $L\to L_c$,
accompanied by anomalous transition behavior.  However, when viewed on an
`anomalous' lengthscale ($O(T^{1/3})$ or $O(T^{1/4})$ or ..., depending on
the form of the potential), those fluctuations remain finite, and one would
presumably observe a rounded maximum of the prefactor (when scaled by the
appropriate temperature-dependent factor) at $L_c$.

The situation is therefore fairly subtle; care must be used to describe the
`transition' in appropriate terms.  Moreover, in perhaps the most important
respect, the transition fails a central test of criticality --- that of the
disappearance of a characteristic fluctuational lengthscale.  In our
problem, even at $L_c$ there is a well-defined characteristic lengthscale,
albeit an anomalous one.

This is strikingly different from a similar-looking (on the surface)
transition described elsewhere~\cite{MS93,DMS94,MS96}.  This transition in
the activation behavior of {\it non-equilibrium\/} systems (e.g., where
detailed balance is absent in the zero-noise dynamics) occurs as the result
of singularities developing~\cite{MS00} in the action.  These singularities
lead in turn to the appearance of {\it caustics\/} in the pattern of
fluctuational trajectories, as in geometric optics.  Here the transition
occurs as a parameter in the zero-noise dynamics is varied.

In these systems, fluctuations about the optimal escape trajectory {\it do
\/} occur at all scales at the critical point, so one cannot simply `cure'
the divergence by including higher-order terms as before.  A {\it de
novo\/} scaling theory~\cite{MS96,MS00} is required.  This theory results
in an array of nontrivial critical exponents that obeying scaling relations
and characterize the divergence (or vanishing) of relevant physical
quantities (including, but not limited to, the rate prefactor).
Consequently, these non-equilibrium systems can justifiably be said to
exhibit true critical behavior at the transition point.

The discussion so far has focused entirely on second-order transitions;
what about first-order?  This possibility has been discussed by several
authors~\cite{Chudnovsky92,KT97,GB97}.  In~slightly more complicated
classical field theories perturbed by spatiotemporal noise, such as a
sixth-degree Ginzburg--Landau model, the nonconstant saddle branch of the
energy functional ${\cal H}[\cdot]$ can in principle cross the constant
saddle branch at a nonzero angle.  This should give rise to a first-order
transition.  So, in the phase plane of these models, the second-order
transition point (in the limited sense discussed in this section) is
presumably the endpoint of a first-order transition curve.

\subsection{Is the quantum tunneling~$\leftrightarrow$~classical activation
crossover for a single degree of freedom identical to the transition in
activation behavior of a classical field on finite intervals?}
\label{subsec:identical}

Mathematically, yes, under the following mapping:

\medskip

\begin{tabular}{ccc}
& $\underline{{\rm Classical} \leftrightarrow {\rm Quantum}}$ &
$\underline{\hbox{Classical }\phi^4}$\\ &&\\ Small parameter & $\hbar$ &
$T$\\ Tunable parameter & $T$ & $L$\\ Periodic in & $\beta\hbar$ & $L$\\
\end{tabular}

\medskip

Here we chose a stochastic Ginzburg-Landau $\phi^4$ model for specificity,
but one can substitute any of the other classical field theories discussed.
For the noisy magnetization dynamics discussed in
Sec.~\ref{subsec:nanomagnets}, it should be remembered that the transition
can occur as either length or field is varied.  One also need not choose
periodic boundary conditions for classical Ginzburg-Landau field theories;
the transition will occur in similar fashion (but with fairly minor
differences as described in Sec.~\ref{sec:finite}) with other types of
boundary conditions.  On the other hand, one is constrained to use boundary
conditions periodic in $\beta\hbar$ in the
quantum~$\leftrightarrow$~classical problem.

This mapping is realized when the quantum~$\leftrightarrow$~classical
transition problem is set up using a Euclidean imaginary-time functional
integral formulation, in a potential where incoherent tunneling dominates.
The classical problem is approached similarly as a real-space path integral
in the limit of weak spatiotemporal noise.

But there is a significant physical difference (leaving aside the obvious
ones) between the two classes of problems.  In the previous section, we
were able to consider --- for the classical problem --- either holding $L$
fixed and lowering $T$, or holding $T$ fixed and moving $L$ through $L_c$.
In the first case one can, in principle, recover the prefactor divergence
shown in Fig.~\ref{fig:prefactor}, while in the second one should observe a
sequence of Arrhenius $\rightarrow$ non-Arrhenius $\rightarrow$ Arrhenius
transition behaviors, with the width of the non-Arrhenius region vanishing
as $T\to 0$.  `Non-Arrhenius' means, as usual, a prefactor $\Gamma_0$ not
determined by~(\ref{eq:gamma0}), and thereby usually aquiring a temperature
dependence.  In order to see this dependence, one would need to repeat the
procedure (fix $T$, move $L$ through the transition) at several different
temperatures.

But in the quantum case, there is no such freedom: here the small
parameter, $\hbar$, cannot be varied.  So a prefactor divergence such as
Fig.~(\ref{fig:prefactor}) cannot be observed even in principle.

In a similar vein, the zero mode arising from the use of periodic boundary
conditions will differ in the two cases.  This mode arises on the
nonconstant (`bounce') side of the transition due to translational symmetry
--- the instanton (or domain wall) separating the two stable solutions can
arise anywhere in space (or imaginary time, in the
quantum~$\leftrightarrow$~classical case).  This leads in turn to a
$T$-dependent (respectively, $\hbar$-dependent) prefactor for {\it all\/}
$L>L_c$ (respectively, $T<T_c$).  

It should be kept in mind that we have made the implicit assumption in the
classical field case that temperature can be lowered to arbitrarily small
values.  For many systems this may not be the case: either transition rates
will become immeasurably low, or the system itself might undergo a (real)
phase change, or new physics may enter in some other way.  Nevertheless,
the ability to vary temperature does distinguish the classical field
transition from the quantum~$\leftrightarrow$~classical one.  In
particular, the magnetic reversal problem
(cf.~Sec.~\ref{subsec:nanomagnets}) presents us with the exciting
possibility of continuously tuning through the transition by varying an
external magnetic field.  With the possibility this presents of
experimentally studying activation closer to the transition than might
otherwise be achieved, we might in this way not only advance our
understanding of stochastic reversal in nanoscale magnets, but also uncover
new information about the quantum~$\leftrightarrow$~classical transition
that the constancy of $\hbar$ would otherwise prevent us from obtaining.

\section{Acknowledgments}

This research was partially supported by the U.S.~National Science
Foundation Grants No.~PHY-0099484 and PHY-0351964.


\end{document}